\documentclass[%
 reprint,
 amsmath,amssymb,
 aps,
pra,
]{revtex4-2}

\usepackage{graphicx}%
\usepackage{dcolumn}%
\usepackage{bm}%
\usepackage{amsmath}
\usepackage{leftindex}
\usepackage{placeins}

\begin{document}

\preprint{APS/123-QED}

\title{Orientation-Dependent Enhanced Ionization in Acetylene Revealed by Ultrafast Cross-Polarized Pulse Pairs}%

\author{S. A. Mohideen$^{1}$}
\author{A. J. Howard$^{2,3}$}%
\author{C. Cheng$^{1,3}$}%
\author{I. Gabalski$^{2,3}$}%
\author{A. M. Ghrist$^{2,3}$}%
\author{E. Weckwerth$^{1,3}$}%
\author{P. H. Bucksbaum$^{1,2,3}$}%
\email{Corresponding author. Email: phbuck@stanford.edu}

\affiliation{$^1$Department of Physics, Stanford University, Stanford, CA, 94305, USA;}
\affiliation{$^2$Department of Applied Physics, Stanford University, Stanford, CA, 94305, USA;}
\affiliation{$^3$Stanford PULSE Institute, SLAC National Accelerator Laboratory, 2575 Sand Hill Road, Menlo Park, CA, 94025, USA}

\date{\today}%

\begin{abstract}
We investigate the orientation dependence of Enhanced Ionization (EI) during strong-field-driven nuclear motion in acetylene (C$_2$H$_2$). 
Here, we both initiate and probe molecular dynamics in acetylene with intense 6-fs cross-polarized pulse pairs, separated by a variable delay.
Following multiple ionization by the first pulse, acetylene undergoes simultaneous elongation of the carbon-carbon and carbon-hydrogen bonds, enabling further ionization by the second pulse and the formation of a very highly charged state, [C$_2$H$_2]^{6+}$.
At small inter-pulse delays ($<$20 fs), this enhancement occurs when the molecule is aligned to the probe pulse.
Conversely, at large delays ($>$40 fs), formation of [C$_2$H$_2]^{6+}$ occurs when the molecule is aligned to the pump pulse.
By analyzing the polarization and time dependence of sequentially ionized [C$_2$H$_2]^{6+}$, we resolve two distinct pathways that both contribute to a large increase in the multiple ionization yield.
This cross-polarized pulse pair scheme uniquely enables selective probing of deeply bound orbitals, providing new insights on orientation-dependent EI in highly charged hydrocarbons.
\end{abstract}

\maketitle

\section{\label{sec:level1}Introduction}

Nuclear dynamics initiated by strong-field ionization have been studied extensively in many diatomic and triatomic species \cite{bocharova_charge_2011, howard_filming_2023, zuo_charge-resonance-enhanced_1995}. 
A subset of these studies have found a characteristic increase in ion yield as a function of molecular geometry and have labeled the phenomenon Enhanced Ionization (EI). 
EI can occur when ionization induces bond elongation, allowing the molecule to reach a transient ``critical'' geometry where the probability of ionizing additional electrons is maximized. 
Prior studies have attributed the increased probability of ionization to the localization of electrons at particular atomic sites and the simultaneous suppression of the potential barrier seen by those sites during strong-field tunneling \cite{zuo_charge-resonance-enhanced_1995, seideman_role_1995, posthumus_dissociative_1995}. Thus far, critical geometries that trigger EI have been identified in various diatomic and triatomic molecules (e.g: H$_2$, D$_2$, I$_2$, H$_2$O, N$_2$O, CO$_2$) \cite{zuo_charge-resonance-enhanced_1995, xu_experimental_2015, howard_filming_2023, seideman_role_1995, bocharova_charge_2011, ueyama_concerted_2005, pavicic_intense-laser-field_2005}. 

Many studies have aimed to investigate how EI manifests in larger polyatomic species (including acetylene) \cite{russakoff_time-dependent_2015, gaire_photo-double-ionization_2014,ohrendorf_dicationic_1990,roither_high_2011,burger_time-resolved_2018}. 
Specifically, experimental efforts have attempted to image the nuclear dynamics in acetylene using strong fields to ionize and subsequently probe the molecule after some time delay \cite{burger_time-resolved_2018,erattupuzha_enhanced_2017, gong_strong-field_2014,  hartmann_ultrafast_2019}.
Some of these studies note the occurrence of EI following the ionization-driven nuclear dynamics in multiply ionized acetylene, specifically elongation of the carbon-hydrogen (CH) bonds in doubly ionized acetylene \cite{burger_time-resolved_2018}. 
Here, we present evidence for the importance of simultaneous carbon-carbon (CC) and carbon-hydrogen (CH) bond elongation en route to EI.

The interpretation of the unexpectedly high ionization yields observed in polyatomic molecules such as CH$_4$, C$_2$H$_4$, C$_3$H$_8$O, C$_4$H$_6$, and C$_6$H$_{14}$ remains controversial \cite{dewitt_concerning_1999,roither_high_2011,mishra_ultrafast_2022}. 
Some studies invoke EI to explain the substantial yields of highly charged ions in these large hydrocarbon molecules \cite{xie_role_2014, roither_high_2011}. 
These theories suggest that tunneling ionization occurs simultaneously in multiple (spatially) parallel bonds due to suppression of the tunneling barrier in each.
This is purportedly responsible for the removal of as many as 12 electrons in large hydrocarbons such as  1,3-butadiene \cite{roither_high_2011}. 
Other experimental and theoretical research suggests unique forms of EI involving the up-shifting of and couplings between more deeply bound molecular orbitals \cite{erattupuzha_enhanced_2017, russakoff_time-dependent_2015}.

Ab-initio studies indicate that the orbital vacancies formed by strong-field ionization in acetylene heavily depend on the orientation of the molecular axis with respect to the polarization direction of the applied field \cite{erattupuzha_enhanced_2017, gong_strong-field_2014, russakoff_time-dependent_2015, xie_electronic_2014}. 
Therefore, some formulations of EI in acetylene rely on these findings to suggest that the enhancement mechanism is greatly orientation-dependent. 
Specifically, these studies demonstrate the significant impact of CH bond stretching in making highly bound orbitals energetically favorable for ionization under strong fields with specific orientations \cite{erattupuzha_enhanced_2017,russakoff_time-dependent_2015}. 
Similar effects have been studied previously in water and proven to greatly affect the EI phenomenon \cite{howard_filming_2023}.

This study introduces polarization-dependent experimental data to augment the current understanding of how strong field ionization can elicit nuclear motion and drive EI in the cationic states of acetylene.  
Probing these states with pairs of cross-polarized pulses reveals previously unseen features of the molecular dynamics. 
Specifically, we uncover polarization- and time-dependent mechanisms by which EI manifest in acetylene, involving the simultaneous elongation of both its CC and CH bonds.

\section{\label{sec:level2}Methods}
The 6-fs 750-nm 1-kHz strong-field pulse pairs that were used to both initiate and probe molecular dynamics in this experiment were generated in a multi-stage ultrafast infrared laser system. 
The first stage consists of a Titanium:sapphire (Ti:sapph) oscillator, followed by a regenerative Ti:sapph chirped pulse amplification system.
The output of this second stage is a 2-mJ pulse of $\sim$40 fs duration and 800-nm center wavelength. 
In the next stage, 450 $\mu$J of the pulse is directed into a hollow-core fiber with a 500~$\mathrm{\mu}$m diameter and a gradient pressure of argon gas (ranging from 0 to 1.3 bar over the 2.5-m length of the fiber) where it is spectrally broadened \cite{robinson_generation_2006}. 
After spectral broadening, the pulse is re-compressed with a total of 16 bounces between two chirped mirror blocks in a V-formation \cite{steinmeyer_new_2001,boyd_nonlinear_2019}. 
The final temporal width of $\sim 6$~fs is characterized by the full width at half maximum (FWHM) of the intensity profile after the re-compression and pulse characterization.
The pulse profiles are characterized with a second-harmonic dispersion scan, in which a spectrometer captures the spectral power of the second-harmonic after the pulse passes through two BK7 wedges of variable insertion and a 10-$\mathrm{\mu}$m thick beta barium borate (BBO) crystal \cite{miranda_simultaneous_2012}.

The collimated beam then enters a Mach-Zehnder interferometer that splits the beam into two arms of equal intensity. 
The probe arm rests on a mechanically actuated delay stage to allow variable pulse separations (between -10 and +90 fs). 
Before recombination, each arm passes through a wire grid polarizer to align the pump polarization along the lab-frame $z$-axis and the probe polarization along the lab-frame $y$-axis (with the propagation direction along the lab-frame $x$-axis). The cross-polarized beams have equal intensities ($14~\mathrm{\mu}\text{J}$, 2$\times$10$^{15}$~W/cm$^2$). 
After re-combination, the output of the interferometer is sampled and sent to a spectrometer which tracks spectral interference to characterize the temporal delays between the pulse pairs \cite{diels_ultrashort_2006}. 
The beam is then sent into an ultra-high vacuum chamber with a background pressure of $6 \times 10^{-10} {\text{ Torr}}$, where it is focused by a spherical metal mirror ($f$=5~cm and focal spot $\sim$10~$\mu$m).
An effusive source of gaseous acetylene passes through the focus at a low enough pressure ($1 \times 10^{-9}~\text{Torr}$) such that, on average, $<$1 molecule is in the focus per pump$/$probe pair.

\begin{figure}[ht]
    \includegraphics[width=\columnwidth]{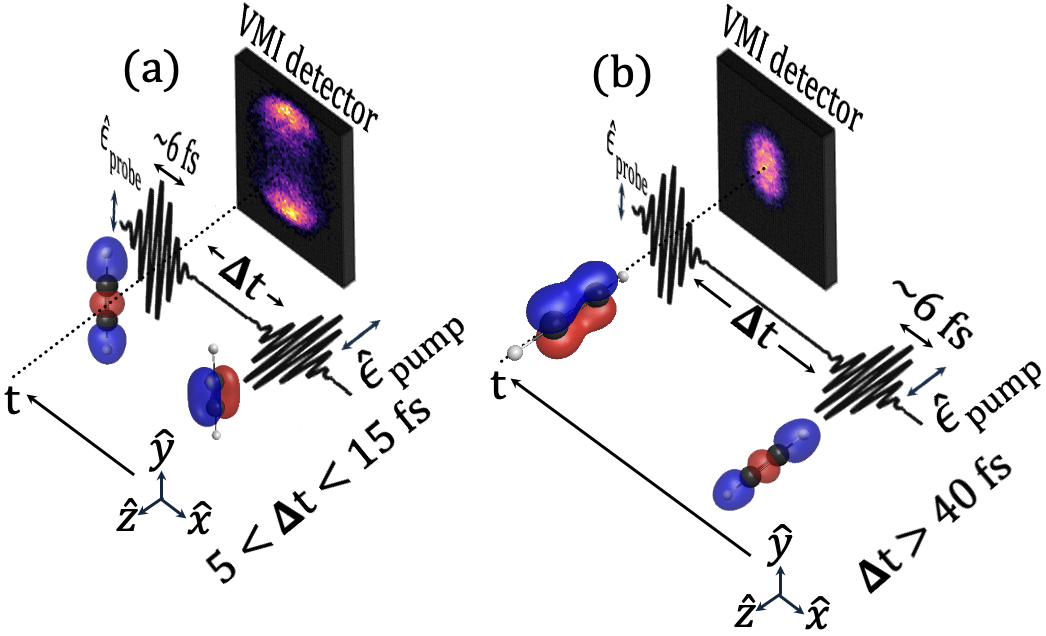}
    \caption{
    Experimental schematic. 
    The two diagrams depict the perpendicular molecular orientations during \textbf{(a)} early and \textbf{(b)} late time delays.
    The pump pulse (polarized along the z-axis) multiply ionizes the molecule. 
    After a delay ($\Delta t$), the probe pulse (polarized along the $y$-axis) further removes electrons,  resulting in a four-body dissociation. 
    The H$^+$/H$^+$/C$^{2+}$/C$^{2+}$ ions are then captured in coincidence and their time of flight as well as their $x$ and $y$ positions on the detector are recorded, allowing the 3D reconstruction of their momentum.
    The orbitals with the highest ionization probability (given the relative angle of the polarization axis to the molecule) are illustrated on each configuration.   
    Panel (a) illustrates the molecular orientation along the $y$-axis at short-time delays ($5 {\text{ fs}}< \Delta t < 15 {\text{ fs}}$).
    Panel (b) illustrates the molecular alignment along the $z$-axis at large inter-pulse delays ($\Delta t > 40$~fs). 
    }
\end{figure}

Figure 1 schematically outlines the experimental setup.
Two cross-polarized 6-fs strong-field pulses encounter an acetylene molecule in the vacuum chamber, at the beam's focus.
The pump pulse, polarized along the $z$-axis, drives initial multiple ionization of the acetylene molecule, primarily forming the dication ([C$_2$H$_2]^{2+}$) but also the trication ([C$_2$H$_2]^{3+}$) and tetracation ([C$_2$H$_2]^{4+}$) in an approximate ratio of 10:2:1.

The multiply charged molecule then undergoes nuclear dynamics on the femtosecond timescale. 
The dynamics persist until a cross-polarized probe pulse (polarized along the $y$-axis) further ionizes the molecule after a variable amount of time (up~to~90~fs), generating the six-times-ionized state, [C$_2$H$_2]^{6+}$.
This highly charged state rapidly Coulomb explodes into four individual ionic fragments (H$^{+}$/H$^{+}$/C$^{2+}$/C$^{2+}$). 
An electric field of strength $\sim$667 V/cm forces the positive fragments towards a high resolution time- and position-sensitive ion detector. 
This detector is comprised of a triple stack of micro-channel plates and a Roentdek hexanode delay-line detector \cite{jagutzki_multiple_2002}. 
After the fragments land in coincidence on the detector, we reconstruct the full three dimensional momenta of each fragment upon dissociation. 
This allows us to deduce the geometry and orientation of each acetylene molecule upon dissociation.

\section{\label{sec:level3} Reconstructing Dynamics from the Coincident Ion Momenta}

Following Coulomb explosion of the six-times-ionized molecule into four ionic fragments (H$^{+}$/H$^{+}$/C$^{2+}$/C$^{2+}$), we reconstruct the full three dimensional momenta ($p_{\hat{x}_\mathrm{m}}$, $p_{\hat{y}_\mathrm{m}}$, and $p_{\hat{z}_\mathrm{m}}$) of each ionic fragment in the lab-frame directly using their detected positions and times-of-flight. 
The lab-frame ion momenta are then used to create a retrieved molecular frame with the following basis vectors: 
\begin{subequations}
\label{Eq_1}
\begin{align}
    \hat{z}_\mathrm{m} & = 
    \left(\frac{ \vec{p}_\mathrm{C_{(1)}} }{|\vec{p}_\mathrm{C_{(1)}}|} + \frac{ \vec{p}_\mathrm{C_{(2)}} }{|\vec{p}_\mathrm{C_{(2)}}|}
    \right) /\hspace{1mm} 2\mathrm{cos}(\beta/2) \label{Eq_1a}\\
    \hat{x}_\mathrm{m} &= \left(\vec{p}_\mathrm{C_{(1)}} \times \vec{p}_\mathrm{C_{(2)}}\right) /\hspace{1mm} |\vec{p}_\mathrm{C_{(1)}}| \hspace{0.25mm} |\vec{p}_\mathrm{C_{(2)}}| \hspace{0.5mm} \mathrm{sin}(\beta) \label{Eq_1b}\\
    \hat{y}_\mathrm{m} &= \hat{z}_\mathrm{m} \times \hat{x}_\mathrm{m} \label{Eq_1c}\\
    \notag\\
    &\hspace{-4mm} \mathrm{where} \hspace{1mm} \beta = \mathrm{arccos}(\vec{p}_\mathrm{C_{(1)}} \cdot \vec{p}_\mathrm{C_{(2)}}) \notag
\end{align}
\end{subequations}
Here, $\hat{z}_\mathrm{m}$ is the bisector of the two C$^{2+}$ momenta, $\hat{x}_\mathrm{m}$ is the vector normal to the plane containing the two C$^{2+}$ momenta, and $\hat{y}_\mathrm{m}$ is the in-plane vector perpendicular to $\hat{z}_\mathrm{m}$.
Fragment momentum, mass, and pump-probe delay ($\Delta t$) are used to create time-resolved plots of total kinetic energy release (KER)--the sum of kinetic energy of every fragment detected in coincidence--upon Coulomb explosion. Figure 2(a) shows time-resolved KER for instances where H$^{+}$, H$^{+}$, C$^{2+}$, and C$^{2+}$ fragments were captured in coincidence. 
The molecules undergo multiple ionization from the pump pulse, after which the probe pulse removes additional electrons to finally reach the six-times-ionized ([C$_{2}$H$_{2}]^{6+}$) state prior to Coulomb explosion.

\begin{figure}[htbp!]
\includegraphics[width=\columnwidth]{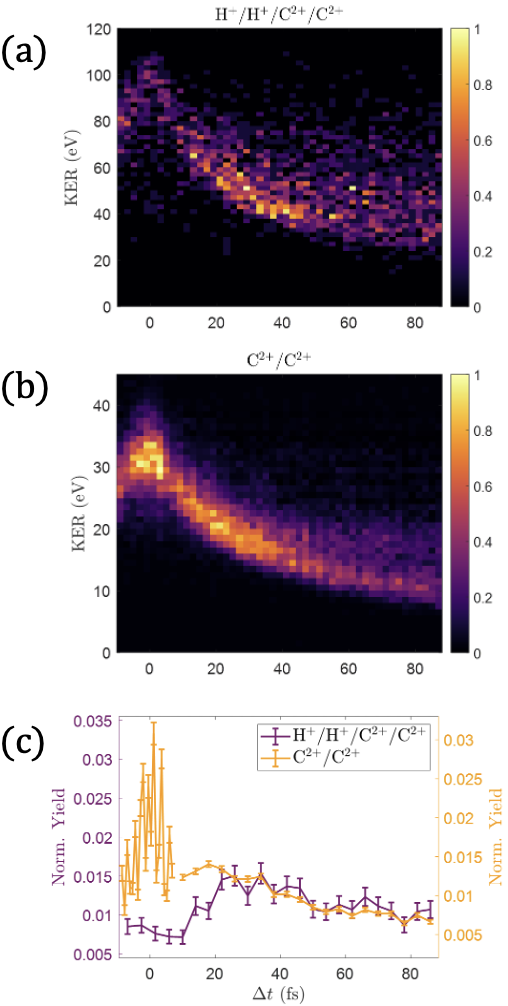}\caption{\label{fig:epsart}\textbf{(a)} Total KER versus pump-probe delay ($\Delta t$) for the four-fold coincidence, H$^{+}$/H$^{+}$/C$^{2+}$/C$^{2+}$, illustrates a monotonic decrease in KER and an increased yield at $\Delta t\approx 20$~to~40~fs. 
\textbf{(b)} Total KER versus $\Delta t$ for the two-fold coincidence, C$^{2+}$/C$^{2+}$, exhibits similar features as seen in panel (a). 
The large peaks in KER near $\Delta t = 0 \text{ fs}$ in panel (b) are due to optical interference of the pump and probe pulses.
\textbf{(c)} Scaled count rate versus $\Delta t$ for the two-fold (yellow) and four-fold (purple) coincidence channels, revealing EI at $\Delta t \approx 20~\text{fs}$. 
All counts are normalized by bin width. The two-fold curve is binned more finely at shorter delays in order to reveal the periodicity of optical interference (2.5~fs).
The large error bars in the four-fold curve reflect the lower statistics of this channel.
}\end{figure}

The time-resolved KER for the two-fold coincidence of C$^{2+}$/C$^{2+}$ in Fig 2(b) exhibits very similar dynamics as the complete four-fold coincidence of H$^{+}$/H$^{+}$/C$^{2+}$/C$^{2+}$ in Fig 2(a). 
Note that the two channels differ for $\Delta t <$~25~fs, due to a technical difficulty in this experiment: at early times, the hydrogens have too high of transverse momentum ($p_{\mathrm{H^+}}^\perp>$~50~$\hbar/a_0$ or KE$_{\mathrm{H^+}}^\perp>$~20 eV) to be captured by our detection methods. 
This is best illustrated by Fig.~1a, where early-time alignment of the molecular axis along the lab-frame $y$-axis results in high transverse momentum relative to the detector plane.
In the interest of comparing Fig.~2(a) and (b), we posit here that the exclusive pathway for producing two C$^{2+}$ ions simultaneously is by formation (and subsequent dissociation) of the six-times-ionized state of the molecule. 
In other words: for two electrons to be removed from the carbon atoms, the hydrogen atoms must also ionize.
This assumption is justified by the fact that the ionization potential (IP) of atomic hydrogen (13.6~eV) is significantly lower than the double IP of atomic carbon (35.6~eV)~\cite{NIST_ASD}.
In the interest of analyzing higher statistics data, we therefore focus our study on the two-fold coincidence channel (Fig 2(a)) to understand the behavior of the six-times-ionized four-fold coincidences. 
In comparing the two- and four-fold channels, note that both KER distributions (see Figs.~2(a) and 2(b)) exhibit an approximate three-fold increase in yield around $\Delta t$~=~25~fs as compared to late times ($\Delta t$~=~80~fs).

\section{Polarization Dependence of EI}
\subsection{Time-Resolved Molecular Orientations}

The temporal region of high ionization yield occurring between 10~and~35~fs in Fig 2(b) is characteristic of EI. 
Figure 3 illustrates the molecular orientations in this highly charged channel as a function of time delay. 
The time-dependent orientation is characterized by 
\begin{equation}
    \theta = \mathrm{arctan}\left[ (\hat{y}_\mathrm{m} \cdot \hat{y}) / (\hat{y}_\mathrm{m} \cdot \hat{z}) \right],
\end{equation}
defined as the angle of the CC bond (approximated by $y_\mathrm{m}$) in the lab-frame $zy$-plane, versus $\Delta t$. 
Figure 3, also further breaks down the angular distribution information into three one-dimensional polar plots of $\theta$, each representing distinct time-delay bins.
These plots identify three primary configurations. 
At large delays ($\Delta t >$~35 fs), the CC bonds are aligned along the $z$-axis, corresponding to the pump polarization. In contrast, at earlier times, near the onset of the enhancement (5~fs~$<\Delta t<$~15~fs), the CC bonds are aligned along the $y$-axis, corresponding to the probe polarization. 
During the peak enhancement region (20~fs~$<\Delta t<$~35~fs), the polarization is largely isotropic, possibly resembling a statistical mixture of the orientations observed at early and late times.

\begin{figure}[ht!]
\includegraphics[width=\columnwidth]{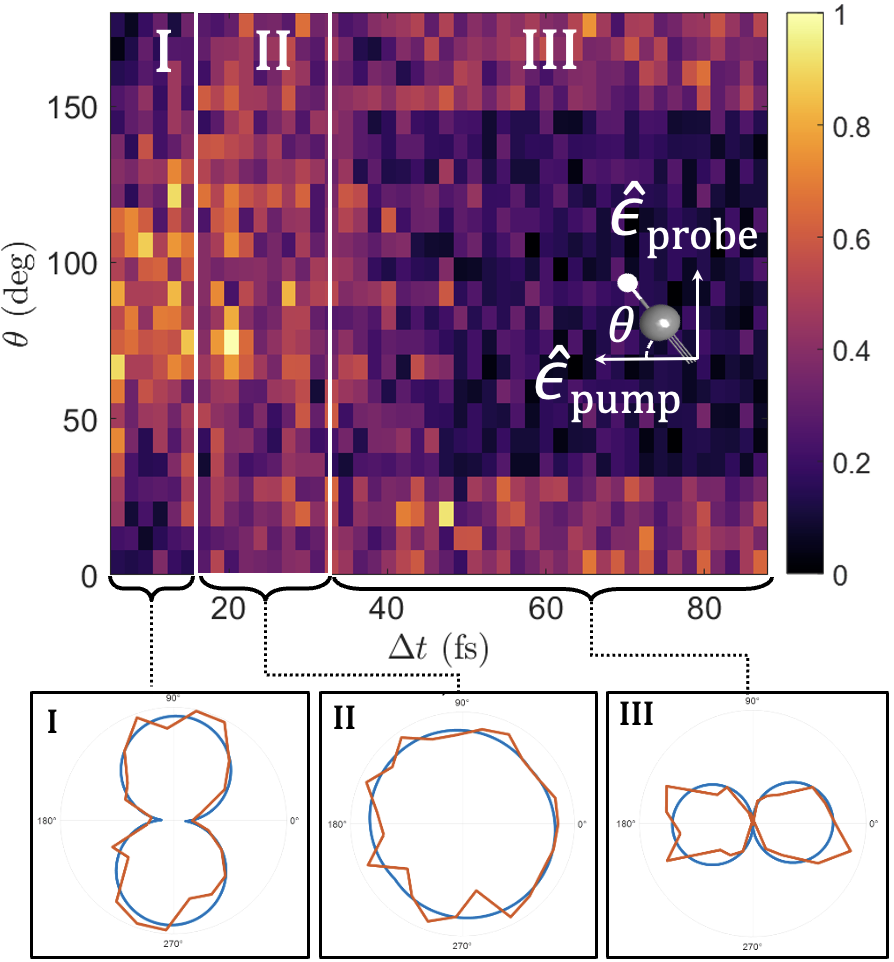}
\caption{\label{fig:Fig3} Time dependent molecular orientations for the two-fold coincidence of C$^{2+}$/C$^{2+}$ ions shown as a two-dimensional plot of $\theta$ versus pump-probe delay ($\Delta t$), where $\theta$ measures the angle between the carbon-carbon bond ($y_\mathrm{m}$) and the lab-frame $z$-axis (probe pulse polarization) in the $zy$-plane.
The key features are extracted into three one-dimensional polar plots of $\theta$, constructed for separate time-delay bins ($5-15$ fs, $15-35$ fs, and $35-90$ fs). 
The raw data plots (orange) are overlaid with a sinusoidal fit (blue) to highlight the dominant shape of the plots in each time bin. 
Region I demonstrates that, at small delays, the molecules are aligned along the $y$-axis (probe pulse direction). 
Region III demonstrates that, at large delays, the molecules are aligned along the $z$-axis (pump pulse direction). 
In Region II, near the maximum enhancement ($\Delta t \approx 25 $ fs), the angular distribution is largely isotropic, possibly an overlap of early and late time orientations.}
\end{figure}
\clearpage

\subsection{Evidence of CC Bond Elongation}

\begin{figure*}[ht!]
\includegraphics[width=\textwidth]{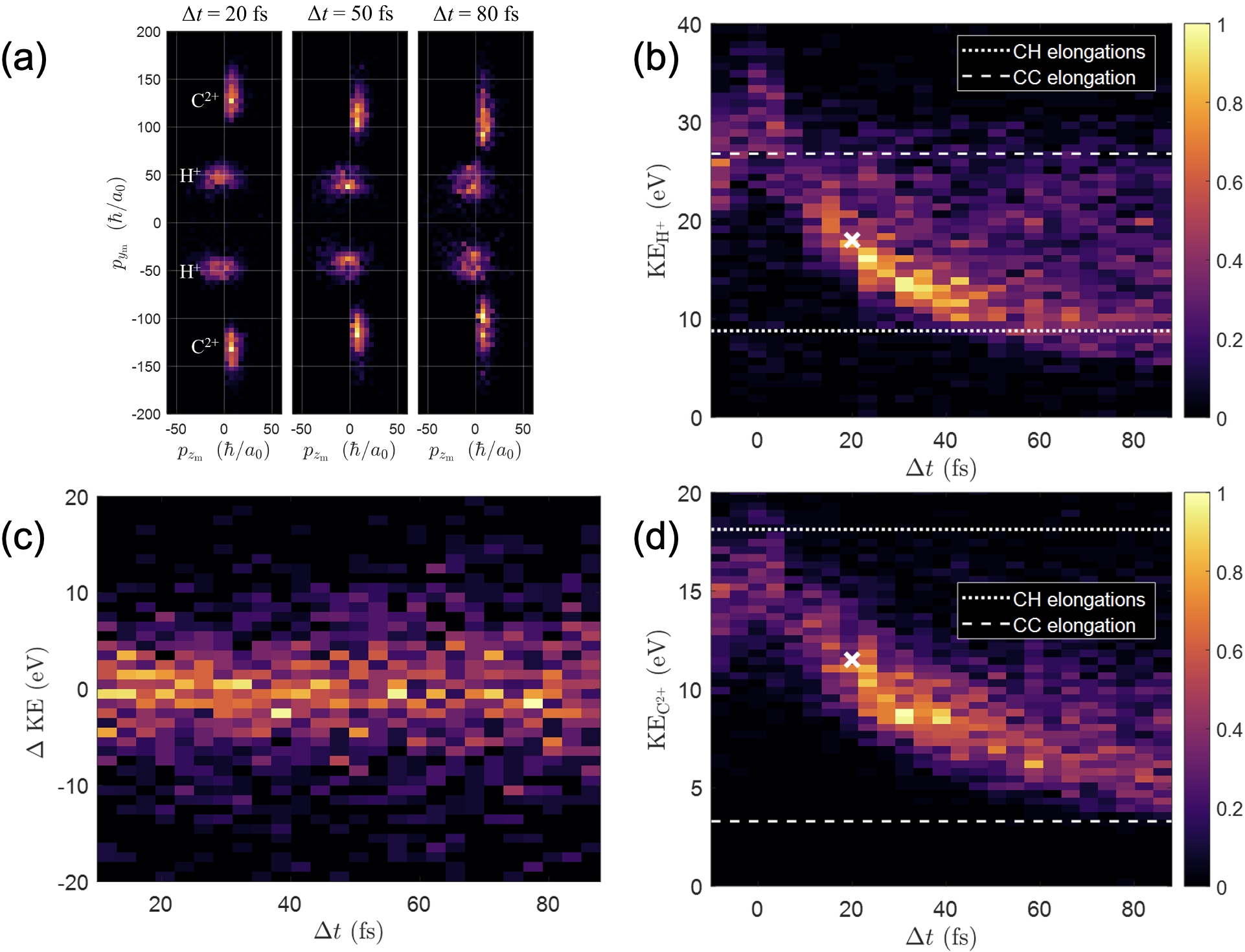}%
\caption{
\textbf{(a)}~The molecular-frame momentum distribution for the four-fold coincidence of H$^+$/H$^+$/C$^{2+}$/C$^{2+}$ shown for three different delays. \textbf{(c)}~The kinetic energy difference ($\Delta KE$) between the two H$^+$ fragments in the same coincidence channel.
\textbf{(b,d)}~The kinetic energy (KE) of (b) H$^+$ and (d) C$^{2+}$ ions in the same coincidence channel as in (a) shown as a function of delay, $\Delta t$.
The dotted lines represent the simulated KE from isolated CH bond elongation (to $r_\mathrm{CH}=20~\text{\AA}$), while the CC bond remains fixed at its equilibrium distance (1.2 \AA). 
The dashed lines indicate the simulated KE for isolated CC bond elongation (to $r_\mathrm{CC} \approx 20\ \text{\AA}$), with the CH bond maintained at its equilibrium length (1.06 \AA).
A white x marks the approximate average KE for each ion at $\Delta t$~=~20~fs, the delay corresponding to the maximum yield in the C$^{2+}$/C$^{2+}$ channel.
}\end{figure*}

In Fig. 4(a), the four-fold fragment momenta are shown in the retrieved molecular frame for $\Delta t $ = 20~fs, 50~fs, and 80~fs. 
The H$^+$ momenta are initially centered around approximately ($p_{\hat{z}_\mathrm{m}}$, $p_{\hat{y}_\mathrm{m}}$) = (0,~$\pm 50$) $\hbar/a_0$, while the C$^{2+}$ momenta are centered around ($p_{\hat{z}\mathrm{m}}$, $p_{\hat{y}\mathrm{m}}$) = (0,~$\pm 125$) $\hbar/a_0$. 
The alignment of all four fragment distributions along $p_{\hat{z}_\mathrm{m}} = 0$ $\hbar/a_0$ at all $\Delta t$ indicates that the molecule does not deviate significantly from linearity, and therefore the dynamics are dominated by linear bond elongations in $r_\mathrm{CH}$ and $r_\mathrm{CC}$ along the central molecular axis.
This finding excludes significant contributions from bending or wagging during the molecular dynamics en route to EI.

Fig. 4(c) depicts $\Delta$KE, the difference in kinetic energy (KE) between the two captured H$^+$ fragments, over $\Delta t$.
The KE of the hydrogen fragments is primarily indicative of their respective CH bond lengths upon Coulomb explosion. 
$\Delta$KE remains tightly distributed about 0 eV for all $\Delta t$.
This identical KE of both hydrogen fragments over delay indicates that both of the CH bonds stretch symmetrically, imparting equal decreases in KE for each of the H$^+$ ions over time.  

Previous studies have attributed EI in acetylene exclusively to elongation of the CH bonds \cite{burger_time-resolved_2018, russakoff_time-dependent_2015}. 
However, our analysis reveals that elongation of both the CC \textit{and} CH bonds contribute to the increased ionization yield observed here. 
In Figure 4(b), the KE for the H$^+$ ions from the four-fold coincidence channel of six-times-ionized acetylene is shown as a function of $\Delta t$. 
The channel exhibits a monotonic decrease in KE from $\sim$30 to 10 eV for increasing pump-probe delay, asymptoting at $\sim$10~eV. 
In Figure 4(d), the delay-dependent KE of C$^{2+}$ illustrates a similar monotonic decrease in KE with an asymptote at $\sim$5~eV. 
The decrease in KE of both species (H$^+$ and C$^{2+}$) is intuitively indicative of simultaneous elongations in the CH and CC bonds. 
Specifically, the decrease in the KE of the hydrogen ions is indicative of CH bond stretching, while the decrease in the KE of the carbon ions points to CC bond stretching. 
An observed decrease in KE from both species in the same channel suggests simultaneous elongations in the CH and CC bonds.

To corroborate our claim of simultaneous CC and CH bond elongation, we performed classical calculations of the Coulomb explosion of the six-times-ionized molecule with two variants of initial conditions: (1) elongation of only the CC bond or (2) symmetric elongation of both CH bonds. 
These models provide theoretical baselines for the asymptotic kinetic energies of each fragment produced during Coulomb explosion. 
Point charges were positioned at the specific atomic locations, and the asymptotic kinetic energies of the individual charges were calculated following evolution of these point charges under simple pairwise Coulomb repulsion according to the ~equation~:
\begin{equation} \label{eq:V_CEI}
    V = \sum_{j \neq i}^N{\frac{q_i q_j}{|\vec{r}_i - \vec{r}_j|}}
\end{equation}
where $q_i$ is the net charge on each fragment, and $\vec{r}_i$ is the three-dimensional position vector of each fragment.
The initial internuclear distances were set according to the equilibrium and asymptotic bond lengths of $r_\mathrm{CH} = 1.06\ \text{\AA}$ and $r_\mathrm{CC} = 20\ \text{\AA}$ or $r_\mathrm{CH} = 20\ \text{\AA}$ and $r_\mathrm{CC} = 1.2\ \text{\AA}$. 
Using this method, we can predict asymptotic kinetic energy values for individual H$^+$ and C$^{2+}$ coincidences at very large interpulse delays, considering isolated elongation of either the CH bonds or the CC bond. 
These asymptotic predictions are overlaid onto Figures 4(b) and 4(d) for comparison.
In Figure 4(b), the calculated asymptotic H$^+$ energy for pure CH elongation is roughly aligned with the observed asymptote of $\sim$10 eV, but the asymptote for pure CC elongation, $\sim$27 eV, is not.
Conversely, in Figure 4(d), the calculated C$^{2+}$ energy for pure CC elongation is roughly aligned with the measured asymptote of $\sim$5 eV, but the asymptote for pure CH elongation, $\sim$19~eV, is not. 
The large discrepancy with this simple model of isolated bond stretching, verifies the necessity of simultaneous elongation of the CH and CC bonds in the formation of six-times-ionized acetylene. 

\clearpage

\subsection{The Enhanced Ionization Mechanism}

\begin{figure}[h]
\includegraphics[width=\columnwidth,trim={0.4cm 0.5cm 0.4cm 0.6cm},clip]{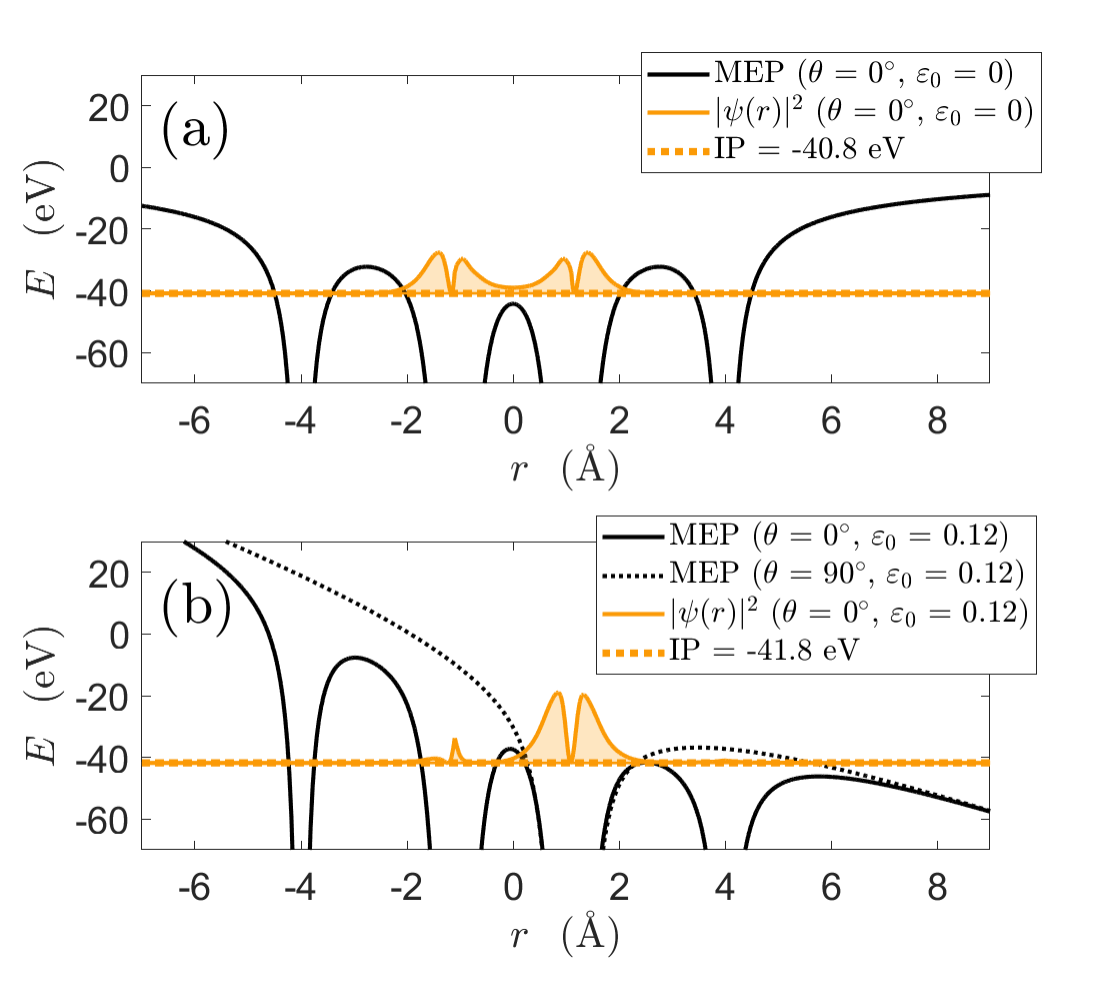}%
\caption{\label{fig:eFig5}
\textbf{(a)} A 1D cut of the field-free molecular electrostatic potential (MEP) of [C$_{2}$H$_{2}]^{5+}$ along the molecular axis, $\theta$=0$^\circ$ (in solid black) for $r_\mathrm{CC}$~=~2.2~$\text{\AA}$ and $r_\mathrm{CH}$~=~2.9~$\text{\AA}$.
A 1D cut of the probability density for the $\sigma$-type HOMO orbital is plotted on the same axis (in shaded orange), sitting atop the ionization potential (IP) (in dotted orange).
\textbf{(b)} Two 1D cuts ($\theta=0^\circ$ and 90$^\circ$ in solid and dotted black, respectively) of the MEP for [C$_{2}$H$_{2}]^{5+}$ distorted by a strong static field of strength 0.12~$E_\mathrm{h}$/$e a_0$.
In each case, the field is directed along the direction of the 1D cut and the nuclear geometry is the same as in panel (a).
A 1D cut of the probability density for the HOMO is again plotted on the same axis (in shaded orange) sitting atop the IP (in dotted orange).}
\end{figure}

The measured alignment of the molecule at early times suggests the dominance of EI in determining the molecular orientation. 
For EI to occur upon the arrival of the probe pulse, the molecular axis must be aligned with the probe polarization.
Figure 3 reveals this tendency ($\theta=\pm 90^\circ$) predominantly at delays just prior to the peak of the enhancement (5-15 fs). 
To elucidate the EI mechanism at early times, we performed a quantum chemistry calculation using GAMESS \cite{barca_recent_2020} to simulate the molecular electrostatic potential (MEP) experienced by an electron in the highest occupied molecular orbital (HOMO) of symmetrically-stretched five-times ionized acetylene distorted by a strong field, resulting in Fig.~5.
Here the MEP and HOMO for the doublet ground state of [C$_2$H$_2$]$^{5+}$ were both generated neglecting exchange interactions using restricted open-shell Hartree-Fock theory with a 6-31G basis set in the absence of any electric field for Fig.~5(a) and in a DC electric field of strength 0.08~$E_\mathrm{h}$/$ea_0$ for Fig.~5(b). 
We utilized a slightly lower DC field strength in this simulation---as compared to the strength of the linear potential gradient (0.12~$E_\mathrm{h}$/$ea_0$) applied in Fig.~5(b)---in order to ensure all electronic orbitals remained bound.

Figure 5 displays a 1D cut of (a) the field-free and (b) the field-distorted MEP for $r_\mathrm{CC}=2.2~\text{\AA}$ and $r_\mathrm{CH}=~2.9~\text{\AA}$ (chosen to approximately reproduce the measured H$^{+}$ and C$^{2+}$ KE at $\Delta t$~=~20~fs, as displayed in Fig.~4(b,d)).
Three separate 1D cuts are shown: one case along the molecular axis ($\theta=0^\circ$) without any applied field (see Fig.~5(a)), another along the molecular axis with an applied field of $\varepsilon_0=0.12 ~E_\mathrm{h}/ea_0$, and one case perpendicular to the molecular axis ($\theta~=~90^\circ$) with the same applied field (see Fig.~5(b)).
We have also displayed the 1D probability density ($|\psi|^2$) of an electron in the $\sigma$-type HOMO of the five-times-ionized molecule, predicting that this orbital would be most likely to ionize given its relatively low IP and $\sigma$ character \cite{erattupuzha_enhanced_2017,russakoff_time-dependent_2015}.

In Fig.~5(a), we see that before any field is applied, the HOMO is distributed largely on the two carbon nuclei, and notably not on either of the hydrogen nuclei.
In Fig.~5(b), we see that under the influence of a strong static field along $\theta=0^\circ$, the HOMO is redistributed primarily on the downhill carbon.
In these conditions, the IP of the HOMO is now just below the downhill ``internal'' potential barrier, and well above the barrier that has formed between the downhill hydrogen and the continuum (the ``external'' tunneling barrier).
Therefore, when a strong field is oriented along the molecular axis of five-times-ionized acetylene, electrons, initially in a $\sigma$-type HOMO, will be localized at the downhill carbon and can tunnel out through an internal (CH) tunneling barrier.
The tunneling probability for this process will be modulated by $r_\mathrm{CH}$, as this distance will determine the height and thickness of both the internal CH barrier and the external tunneling barrier; therefore this effect will only manifest for a critical distance in $r_\mathrm{CH}$.
A similar critical distance may occur in $r_\mathrm{CC}$, as too short or too long of CC distances may preclude the localization of electron density on the downhill carbon.

The required orientation for this proposed EI mechanism is best demonstrated by the comparing the cases of $\theta = 0^\circ$ and 90$^{\circ}$ in Fig.~5(b).
When $\theta$~=~$0^\circ$, the tunneling barrier is significantly thinner (and shorter) than when $\theta$~=~90$^\circ$.
The critical geometry for this enhancement is estimated at $r_\mathrm{CC}$~=~2.2~$\text{\AA}$ and $r_\mathrm{CH}$~=~2.9~$\text{\AA}$.
This geometry was found to best replicate the measured KE of H$^+$ and C$^{2+}$ at the delay of maximum enhancement: $\Delta t$~=~20 fs (see Fig.~4(c) and (d)), assuming the KE of each ion is determined solely by the Coulomb repulsion captured in Eq.~(3).
Curiously, a similar mechanism for EI in the sequential formation of [C$_2$H$_2]^{4+}$ with strong-field pulse pairs was previously measured to peak near $\Delta t = 20 $ fs, and attributed to a critical CH bond length of $r_\mathrm{CH} = 2.0\ \text{\AA}$ during deprotonation of the dication \cite{burger_time-resolved_2018}.
Here we demonstrate complementary evidence for simultaneous CC and CH elongation during the sequential formation of [C$_2$H$_2]^{6+}$.

The late delay data reveal an additional contribution from pump-aligned molecules that undergo simultaneous CC and CH elongation. 
However, the mechanism for enhanced ionization by the probe when the molecular axis is perpendicular is currently unknown.
It is possible that, for late delays (and consequently large values of $r_\mathrm{CC}$ and $r_\mathrm{CH}$), the orbital energies shift to favor ionization from $\pi$ orbitals, rather than EI from $\sigma$ orbitals \cite{erattupuzha_enhanced_2017, doblhoff-dier_theoretical_2016}, resulting in a perpendicularly aligned ensemble.
Regardless of their provenance, the two distinct alignment distributions, at early and late times, converge near 30 fs, as the appearance of pump-aligned molecules coincides with the disappearance of probe-aligned molecules. 
This is illustrated in Region II of Figure 3, where it seems that the overlapping polarization channels create a uniquely isotropic region, resulting in coincidences distributed across all orientations from $0^\circ$ to $180^\circ$. 

\section{Conclusion}
In this experiment, we have measured the  ionization-driven molecular dynamics that elicit EI forming [C$_2$H$_2]^{6+}$.
By sequentially ionizing with strong field pulse pairs, and enforcing that the polarization of the pulses was perpendicular between sequential steps, we were able to generate highly charged states of the acetylene cation, removing as many as 6 electrons. 
Time-resolved coincident fragment momentum imaging allowed us to deduce the molecular geometry and orientation of the molecule, revealing previously unseen instances of simultaneous CC and CH elongation and dramatic orientation-dependent ionization yields.
We identified a mechanism by which simultaneous CC and CH elongation may drive enhanced ionization forming [C$_2$H$_2$]$^{6+}$ at a critical geometry of $r_\mathrm{CC}$~=~2.2~$\text{\AA}$ and $r_\mathrm{CH}$~=~2.9~$\text{\AA}$.
Our work also identified a yet-unexplained contributing alignment distribution at late time delays, wherein [C$_2$H$_2]^{6+}$ is created by a strong field polarized perpendicular to the molecular axis.

In summary, we have uncovered at least two distinct fragmentation pathways in multiply-ionized acetylene that elicit the formation of very high charge states, induced by sequentially ionizing with cross-polarized pulse pairs.
These findings point to the importance of controlling the polarization during strong-field ionization experiments and its effect on the transient electronic states and final ion formation. 
By studying these polarization-dependent dynamics, we not only elicit unseen behavior but also gain deeper insight into how strong-field phenomena such as enhanced ionization manifest.
Our work suggests using the polarization-dependence of strong-field ionization as a mechanism to explore and control molecular ionization and ionic fragmentation processes.
This, in turn, paves the way for new applications of strong field ionization as a versatile tool in molecular imaging, especially as applied to more complex hydrocarbon systems.

\vspace{0.5cm} \section{Acknowledgement}
A.J.H., C.C., I.G., A.M.G., E.W., and P.H.B. were supported by the National Science Foundation. A.J.H. was additionally supported under a Stanford Graduate Fellowship as the 2019 Albion Walter Hewlett Fellow. A.M.G. was funded partially by LCLS Laser Science. A.M.G. was additionally supported by an NSF Graduate Research Fellowship.

\section{References}
\bibliographystyle{apsrev4-2}  %
\bibliography{ref}

\end{document}